\documentclass[prl,twocolumn,aps,superscriptaddress]{revtex4-1}
\usepackage [dvips]{graphicx}
\usepackage{amsmath}
\usepackage{mathrsfs}
\usepackage{color}

\newcommand{\dd}{\text{d}}

\newcommand{\p}{\partial}

\newcommand{\br}{\text{\bf r}}

\newcommand{\eps}{\varepsilon}

\newcommand{\bo}{\text{\bf 0}}

\newcommand{\bu}{\text{\bf u}}

\newcommand{\bnabla}{\boldsymbol{\nabla}}

\usepackage[percent]{overpic}

\begin{document}

\title{Active hard-spheres in infinitely many dimensions}

\author{Thibaut Arnoulx de Pirey}
\affiliation{Université de Paris, Laboratoire Mati\`ere et Syst\`emes Complexes (MSC), UMR 7057 CNRS, F-75205 Paris, France}

\author{Gustavo Lozano}
\affiliation{Departmento de F\'isica, Facultad de Ciencias Exactas y Naturales,
Universidad de Buenos Aires, Ciudad Universitaria, Pabell\'on I, 1428 Buenos Aires, Argentina
}

\author{Fr\'ed\'eric van Wijland}
\affiliation{Université de Paris, Laboratoire Mati\`ere et Syst\`emes Complexes (MSC), UMR 7057 CNRS, F-75205 Paris, France}


\begin{abstract}
Few equilibrium --even less so nonequilibrium-- statistical-mechanical models with continuous degrees of freedom can be solved exactly. Classical hard-spheres in infinitely many space dimensions are a notable exception. We show that even without resorting to a Boltzmann distribution, dimensionality is a powerful organizing device to explore the stationary properties of active hard-spheres evolving far from equilibrium.  In infinite dimensions, we compute exactly the stationary state properties that govern and characterize the collective 
behavior of active hard-spheres: the structure factor and the equation of state for the pressure. In turn, this allows us to account for motility-induced phase-separation. Finally, we determine the crowding density at which the effective propulsion of a particle vanishes. 
\end{abstract}

\maketitle
Understanding the collective behavior of simple liquids has  been a fundamental statistical mechanical challenge since its early days~\cite{hansen1990theory}. The absence of a well-defined and versatile approximation method able to capture collective effects in liquids has led to the development of a branch in its own right: the art of elaborating approximations leading to correlations in fluids is almost as old as statistical mechanics itself~\cite{mayermayer, doi:10.1063/1.1749657,PhysRev.110.1,VANLEEUWEN1959792}. It is only in the mid-eighties that  Frisch, Rivier and Wyler~\cite{PhysRevLett.54.2061} were able to devise a {\it bona fide}  mean-field approximation.  The latter takes the form of a controled large dimensionality limit in which they could derive, among other thermodynamical properties, an exact equation of state for classical hard-spheres. The physical price to pay by going to large space dimensions is heftily compensated by the mathematical gain: not only the equation of state \cite{PhysRevLett.54.2061,PhysRevA.36.2422} but also thermodynamic quantities, such as the entropy~\cite{PhysRevE.60.2942} and even transport coefficients inferred from the collision dynamics~\cite{PhysRevA.37.4351} can be determined exactly. Perhaps more importantly, the greatest insight is to be found in the pair-correlation function in that it, alone, controls the spatial organization of the fluid~\cite{PhysRevA.35.4696}, and can thus be used as an educated starting point for density functional approaches~\cite{doi:10.1080/00018737900101365} (see \cite{10.1007/3-540-45043-2_11} for a recent overview).

The realization that classical infinite-dimensional hard-spheres lent themselves to analytical treatment, especially regarding the determination of entropy, laid the ground for the idea that they could also be used to investigate metastability issues (understood in terms of free energy minima)~\cite{Parisi_2006, RevModPhys.82.789, Kurchan_2012}. They have thus become the workhorse of the theory of jamming and of the static approach to glasses. More recent inroads into dynamical behavior~\cite{PhysRevLett.104.255704,PhysRevE.81.041502,PhysRevLett.116.015902, kurchan2016statics} address relaxation properties, including with nonequilibrium evolutions~\cite{Agoritsas_2019}. For some of these glassy-behavior-related questions, the high-dimensionality comes with its own share of hotly debated issues as to what exactly survives finite dimensions~\cite{PhysRevLett.120.225501}. 
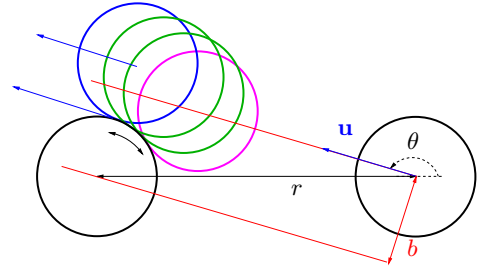
\begin{figure}

\begin{overpic}[totalheight=3.5cm]{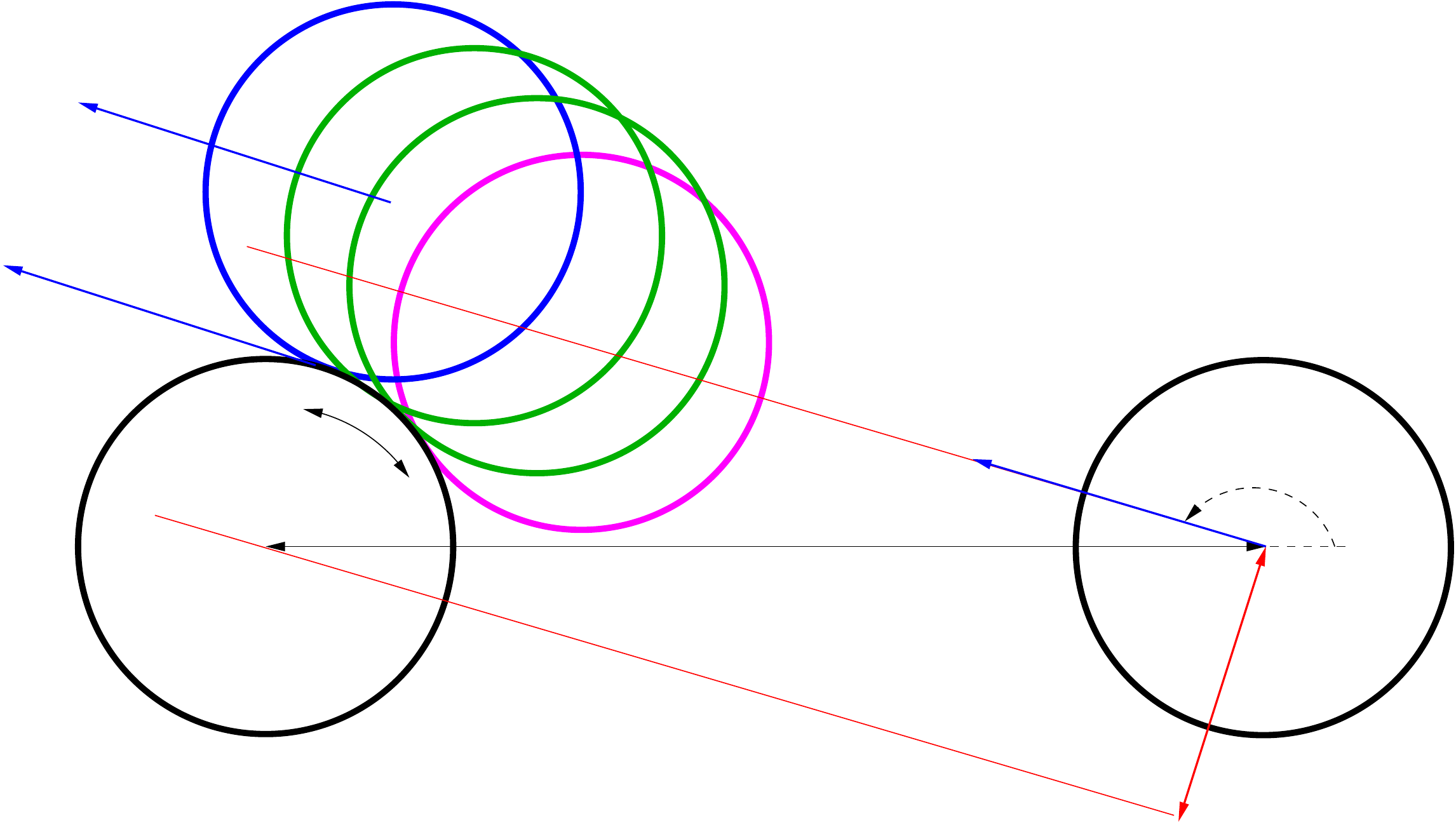}

 \put (85,2) {\textcolor{red}{$b$}}
 \put (70,28) {\textcolor{blue}{$\bu$}}
 \put (85,25) {$\theta$}
 \put (60,15){$r$}
\end{overpic}
  \caption{A collision of an active hard-sphere (black, rightmost) with diameter $\sigma$ and impact parameter $b=r\sin\theta<\sigma$ (and $\cos\theta<0$) onto a pinned (black, leftmost) one. The incoming particle with direction $\bu$ hits the target sphere (at the magenta position) and then skids around by occupying the sequence of green positions. It eventually takes off at the blue position when its orientation $\bu$ becomes tangent to the target sphere. No tumble can occur over the typical skidding distances considered here, which are of order $\sigma/\sqrt{d}$.}\label{fig:collision}
\end{figure}
A  pivotal starting point common to all static approaches is the celebrated equilibrium Boltzmann weight. In stark contrast, no such shortcut exists for the stationary properties of active matter systems and it is thus no surprise that a many-body exactly solvable model of particles interacting with pairwise forces has so far remained elusive. In active systems, the motion of the individual particles requires a net consumption of energy taken from the environment~\cite{ramaswamy-2010, marchetti-2013,fodor2018statistical}. Breaking the delicate balance between dissipation and injection of energy at the particle level inevitably drives even the simplest versions of such interacting particle systems away from equilibrium. Among microscopic models ubiquitous in the active matter literature, the simplest ones involve overdamped dynamics in the presence of a self-propulsion force the statistics of which strongly deviates from the Gaussian white noise familiar in equilibrium. For such systems, even with short-range repulsive interactions, the possibility of a phase-separation into a coexisting dense phase and a dilute one is a direct consequence of the genuine nonequilibrium character of the dynamics. This so-called Motility-Induced Phase Separation (MIPS)  occurs when the typical run length due to self-propulsion notably exceeds the range of repulsive interactions. MIPS is a phenomenon that has received considerable attention~\cite{PhysRevLett.100.218103,PhysRevLett.108.235702,PhysRevLett.110.055701} as it is probably the simplest activity-driven emerging collective phenomenon. Understanding collective behavior in active matter combines the hurdles of strongly correlated liquids with those of nonequilibrium physics. Our purpose is to show how working in infinite dimension allows us to overcome both, and to eventually bridge the microscopic behavior to the macroscopics. In this letter, we begin by defining the proper infinite-dimensional scalings of the model parameters so as to maintain a competition between activity and repulsive pairwise interactions leading to a complex spatial organization. We then solve the two-body problem and use our result to explain how working in large dimension allows us to truncate the hierarchy of correlations to second order.  Relevant physical quantities are then explicitly derived. The effective propulsion velocity~\cite{bialke2013microscopic} is shown to vanish linearly at a crowding density which we identify. The equation of state~\cite{PhysRevLett.114.198301} for the homogenous phase exhibits a regime of negative compressibility that signals the MIPS spinodal, the shape of which is also found exactly.

To carry out this program, our starting point for the dynamics of each particle is an overdamped equation of motion for its position $\br_i(t)$ 
\begin{equation}\label{eq:eqmo}
\frac{\dd\br_i}{\dd t}=-\sum_{j\neq i}\bnabla_{\br_i} V(\br_i-\br_j)+v_0\bu_i
\end{equation}
where the particle's mobility has been set to unity for convenience (without loss of generality), $V(\br)$ is the interaction potential between two particles, $v_0$ is a self-propulsion velocity scale, while $\bu_i$ is a random orientation vector. A variety of models enter this schematic description: for the Run-and-Tumble particles (RTP) we consider here, $\bu_i$ is a unit vector that picks a random direction at rate $\tau^{-1}$ (but, as we discuss in \cite{SM}, our conclusions extend to active Brownian~\cite{cates2015motility} and active Ornstein-Uhlenbeck~\cite{PhysRevE.90.012111} particles). Throughout, the potential $V$ we have in mind is a smooth repulsive potential of the form $V(r)=V_0\exp\left[-\frac{d(r-\sigma)}{\sigma \eps}\right]$ where the $d$ factor keeps it short ranged in the large-dimensional limit~\cite{maimbourg2016approximate}, and where $\eps\to 0^+$ further allows us to take a hard-sphere limit of diameter $\sigma$. The run length between two tumbles is  $\ell=v_0\tau$ and the particle density $\rho$ are the other two dimensionful quantities entering our problem. For noninteracting RTPs the diffusion constant is $\frac{v_0^2 \tau}{d}$ and we choose, as $d\to\infty$, to keep it fixed.
We choose to work at fixed persistence time $\tau$ which leads to keeping $\hat{v}_0=v_0/\sqrt{d}$ fixed. The limit of interest is thus one of a highly ballistic nature where $\ell=\hat{v}_0\tau\sqrt{d}\gg\sigma$ ({\it i.e.} of very large persistence length to particle size ratio $v_0\tau/\sigma$). While other scalings maintaining the nonequilibrium nature of the dynamics are possible (see \cite{SM}), this is  the only one consistent with the emergence of a collective effect such as MIPS. By contrast, the equilibrium limit, while keeping the diffusion constant fixed as well, requires to work at a persistence length vanishingly small with respect to any other relevant scale. Sending $d\to\infty$ first and then $\tau\to 0$ does not allow us here to recover the equilibrium phenomenology. As in \cite{PhysRevLett.54.2061, Parisi_2006,Kurchan_2012,Charbonneau-2017}, we work at density scales such that $\rho V_d(\sigma)\sim O(d)$, so that a given particle typically has $d$ neighbors (here $V_d(\sigma)$ is the exclusion volume of a particle), hence leaving room for nontrivial collective behavior. Therefore, potential gradients are also endowed with a characteristic scale, as we show now. During a collision event between two particles, their relative velocity along the direction of the collision vanishes. The latter features three contributions. The first one accounts for self-propulsion and is of order $v_0/\sqrt{d}$ due to the randomization of the $\bu_i$'s. The second is the two-particle direct interaction of order $\p_r V$. And the third one contains collisions with the rest of the particles: it is a sum over $d$ random contributions (that, for now, we assume to be weakly correlated), each of them being of order  $\p_r V/\sqrt{d}$, hence a global contribution of order $\p_r V$ as well. Altogether we thus expect that $\p_r V$ is of order $\hat{v}_0$.

Let's now discuss the picture that emerges at $d\gg 1$ for just two particles, which amounts to considering the motion of the relative particle with orientation $\bu=\bu_2-\bu_1$ around a fixed spherical obstacle. The impact parameter is given by $b=r\sin\theta $ ($r=||\br||$) as depicted in Fig.~\ref{fig:collision}, but we anticipate that the typical values of interest for $\theta$ are such that $\cos^2\theta\sim d^{-1}$ due to the randomization of $\bu$. The relative motion of an incoming particle a distance $r=\sigma+\delta r$ away from this spherical obstacle is unaffected by the obstacle unless $\delta r/\sigma=O(1/d)$. Indeed if $\delta r/\sigma=O(1)$ a collision event can occur iff $\cos\theta=O(1)$, which is exponentially rare in $d$. When a flip does occur $\delta r/\sigma$ will remain at least of $O(1)$ so that the particle typically misses again the obstacle. Down to these scales, the obstacle is invisible and the particle undergoes a free run-and-tumble motion. This means the density is uniform up to distances $\delta r/\sigma\sim O(1)$. If, however, $\delta r/\sigma$ becomes $O(d^{-1})$ the probability that $\bu$ points towards the obstacle is not negligible anymore so that collision events potentially shape a nontrivial density profile around the obstacle over a scale $\delta r\sim\sigma/d$. We justify this by computing $g_0(\bo,\br;\bu_1,\bu_2)$, the two-point function of the two-body problem for having a particle at $\bo$ with orientation $\bu_1$ and a particle at $\br$ with orientation $\bu_2$. The equation for $g_0$ reads:
\begin{equation}
\label{eq:two_body}
 -v_0 \left(\bu_2 - \bu_1\right)\cdot\bnabla_\br g_0 + 2 \bnabla_\br \cdot \left(g_0\bnabla_\br V(\br) \right)+{\mathscr R}g_0 =0
\end{equation}
where ${\mathscr R}$ is a linear operator acting on $g_0$ and accounting for the dynamics of $\bu_1$ and $\bu_2$ which occurs at a rate $1/\tau$. 
A stimulating inspiration for the solution of Eq.~\eqref{eq:two_body} in the hard-sphere limit comes from the one-dimensional case of two particles on a ring~\cite{PhysRevLett.116.218101}, or of one particle on a finite interval~\cite{malakar2018steady}. In this limit, we can show~\cite{SM} that $g_0(\bo,\br;\bu_1,\bu_2)$ takes the form
\begin{equation}\label{eq:ansaztg}
g_0(\bo,\br;\bu_1,\bu_2)=f(\br;\bu_1,\bu_2)\theta(r-\sigma)+\Gamma(\hat{\br};\bu_1,\bu_2)\delta\left(\frac{r-\sigma}{\sigma}\right)
\end{equation}
where $f$ and $\Gamma$ are  functions yet to be determined, with $\Gamma\neq 0$ only for colliding particles with $(\bu_2-\bu_1)\cdot\br<0$. The extra $\delta$ contribution in Eq.~\eqref{eq:ansaztg} expresses that when a particle collides on another, it skids along at contact for a finite amount of time as depicted in Fig.~\ref{fig:collision}. The regular part $f$ of the profile satisfies:
\begin{equation}\label{eq:regular}
-v_0 \left(\bu_2 - \bu_1\right)\cdot\bnabla_\br f + ({\mathscr R}f) = 0
\end{equation}
while singular part $\Gamma$ is a solution of
\begin{equation}\label{eq:singular}
-v_0(\bu_2-\bu_1)\cdot\left[f(\sigma \hat{\br};\bu_1,\bu_2)\hat{\br} +\nabla_{\hat{\br}}\Gamma-(d-1)\Gamma\hat{\br}\right] + ({\mathscr R}\Gamma) = 0
\end{equation}
This equation expresses the flux balance of incoming particles on the obstacle with those leaving in the course of their skidding around. Given the scale separation between $\sigma$ and and the run length $v_0\tau$, the contributions involving ${\mathscr R}$ can safely be discarded in the $d\gg 1$ limit in Eqs~\eqref{eq:regular} and \eqref{eq:singular}. This allows us to obtain an exact expression for the functions $f$ and $\Gamma$. Denoting by $\theta$ the angle between $\hat{\br}$ and $(\bu_2-\bu_1)$, we obtain:
\begin{equation}
\begin{split}
\label{eq:two_point}
g_0(\bo,\br;\bu_1,\bu_2) = \, & \Theta(r - \sigma)\left[1 - \Theta(\cos\theta)\Theta(\sigma - r \sin\theta)\right] \\ & + \Theta(-\cos\theta)\delta\left(\frac{d(r-\sigma)}{\sigma}\right)
\end{split}
\end{equation}
For colliding particles ($\cos\theta < 0$), there is an accumulation at contact expressed by a delta peak. Since flipping while skidding does not occur in the infinite dimensional limit, there is a depletion of particles away from $r=\sigma$ (hence the conditions $\cos\theta > 0$ and $\sigma - r \sin\theta > 0$ in the regular part). In practice, this depletion is felt over distances $r - \sigma = {O}(\sigma/d)$ ( since $1-\sin\theta\sim 1/d$) and thus bears no effect beyond these scales. In arbitrary dimension, the dimensionless function $\Gamma$ would depend on the ratio $v_0\tau/\sigma$. As $d\gg 1$ this ratio goes to infinity and our final result for $g_0$ is indeed  independent of the dynamical parameters $v_0$ and $\tau$. The spatial distribution function eventually reads
\begin{equation}
g_0(r)=\frac{1}{\Omega_d^2}\int_{\bu_1,\bu_2}\!\!\!\!\!\!\!\!\!\!\!\!g_0(\bo,\br;\bu_1,\bu_2)=\theta(r-\sigma)\left(1+\frac{\sigma}{2}\delta(d(r-\sigma))\right)
\end{equation}
where $\Omega_d$ is solid angle in $d$ dimensions. In the hard-sphere limit, products of the type $g_0(\bo,\br;\bu_1,\bu_2)\bnabla_\br V(\br)$, which are found {\it e.g.} in the virial formula for pressure, also converge to a well-defined distribution. From Eq.~\eqref{eq:two_body}, we show (see \cite{SM}) that 
\begin{equation}\label{eq:regul-2}
\begin{split}
& \lim_{\text{hard sphere}} \int_\sigma^{+\infty} \dd r g_0(\bo,\br;\bu_1,\bu_2)\partial_r V(r) \\ & = \frac{v_0}{2}\left[\left(\bu_2-\bu_1\right).\hat{r}\right] \Gamma(\hat{\br};\bu_1,\bu_2)
\end{split}
\end{equation}
The typical scaling of potential gradients $\partial_r V(r) \sim \hat{v}_0$ discussed earlier is now confirmed. 

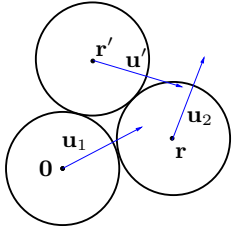
\begin{figure}
\begin{overpic}[totalheight=3cm]{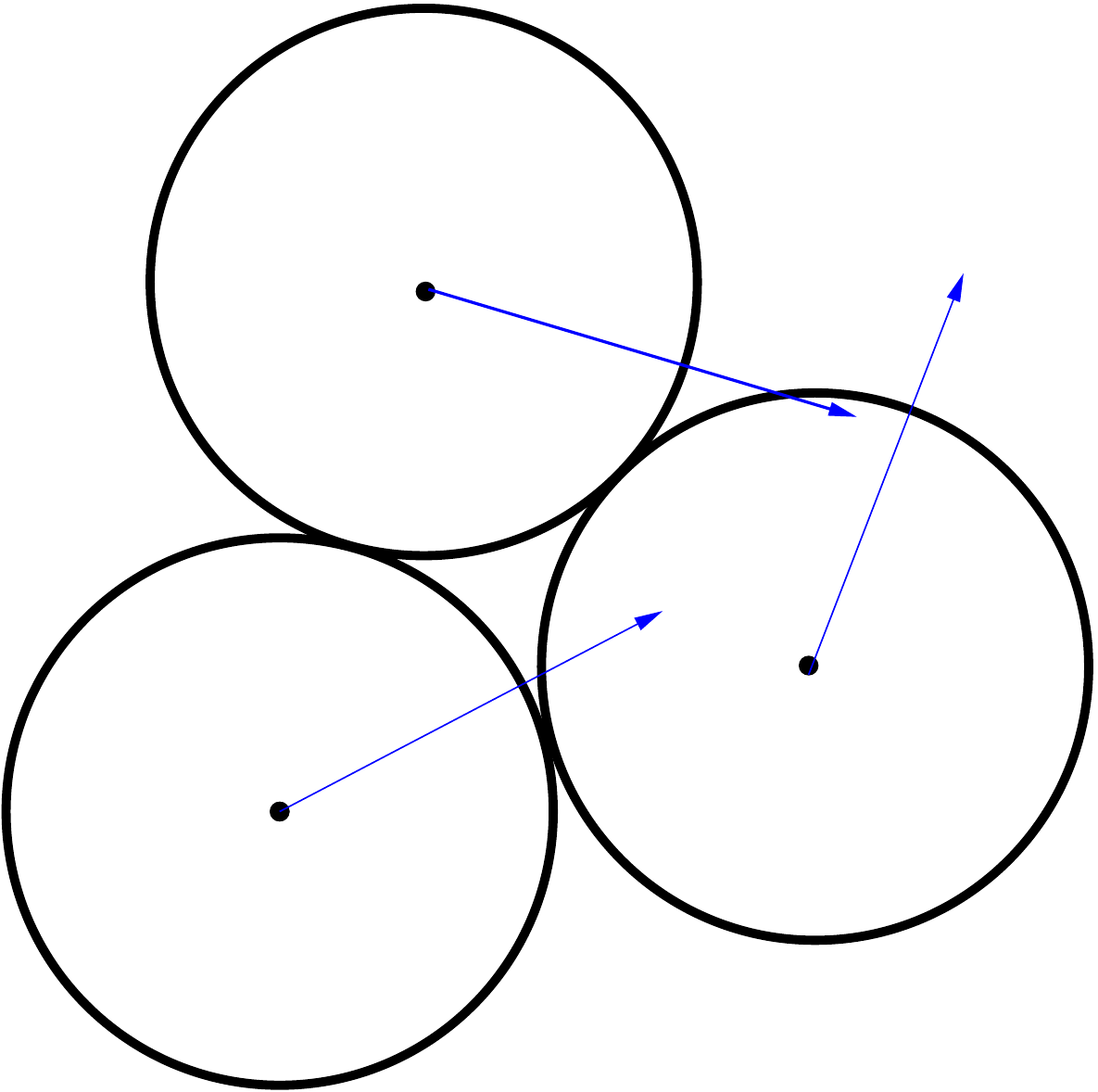}
\put (15,22) {$\bo$}
\put (75,30) {$\br$}
\put (40,75) {$\br'$}
\put (25,35) {$\bu_1$}
\put (80,46) {$\bu_2$}
\put (53,70) {$\bu'$}
\end{overpic}
\caption{Three interacting particles at positions $\bo$, $\br$, $\br'$ with orientations $\bu_1$, $\bu_2$, $\bu'$ forming a loop of contacts, thus being in an unlikely  spatial configuration as $d\to\infty$. \label{fig:tree-unlike}}
\end{figure}

We are now in position to study the $N$-body problem. In the thermodynamic limit, we must deal with the infinite hierarchy of correlation functions inferred from the dynamics. We now sketch the argument that allows us to solve this hierarchy exactly in the $d\gg 1$ limit. This will lead us to conclude that the $N$-body two-point function $g^{(2)}$ actually reduces to $g_0$ determined in Eq.~\eqref{eq:two_point}. The second equation of the hierarchy is given by:
\begin{equation}
\label{eq:hierarchy_second}
\begin{split}
& -v_0 \left( \bu_2 - \bu_1 \right) . \bnabla_{\br} g^{(2)} + ({\mathscr R}g^{(2)}) + 2 \bnabla_{\br}.\left(g^{(2)}\bnabla_{\br}V(\br)\right) \\ & + \rho \bnabla_\br . \left\{ \int \frac{\dd \bu'}{\Omega_d}\dd \br' \, \left( g^{(3)}(\bo,\br,\br'; \bu_1,\bu_2,\bu') \right.\right. \\ & \left.\left. + g^{(3)}(\bo,\br,\br'; - \bu_1,- \bu_2,\bu') \right) \bnabla_{\br'}V(\br') \right\} = 0
\end{split}
\end{equation}
and solving it requires, as usual, the knowledge of $g^{(3)}$.  Assuming a truncation of the hierarchy at the level of the equation for $g^{(3)}$ itself, we show that the resulting equation for $g^{(2)}$ is that of the two-body system. This is at the basis of the systematic proof presented in \cite{SM}. The truncated equation for $g^{(3)}(\bo,\br,\br'; \bu_1,\bu_2,\bu')$ reads
\begin{equation}
\label{eq:hierarchy_third}
\begin{split}
& -v_0\left(\bu_2 - \bu_1 \right). \bnabla_{\br}g^{(3)} - v_0\left(\bu' - \bu_1 \right). \bnabla_{\br'}g^{(3)} + ({\mathscr R}g^{(3)}) \\ & + \bnabla_{\br}.\left[g^{(3)}\left(2 \bnabla_{\br}V(\br) + \bnabla_{\br'}V(\br') + \bnabla_{\br}V(\br-\br')\right)\right] \\ & + \bnabla_{\br'}.\left[g^{(3)}\left(2 \bnabla_{\br'}V(\br') + \bnabla_{\br}V(\br) + \bnabla_{\br'}V(\br'-\br)\right)\right]  = 0
\end{split}
\end{equation}
This equation has the solution $g^{(3)}(\bo,\br,\br';\bu_1,\bu_2,\bu') = g_0(\bo,\br;\bu_1,\bu_2)g_0(\bo,\br';\bu_1,\bu')g_0(\br,\br';\bu_2,\bu')$ up to $O(d^{-1/2})$ corrections. This structure is identical to the one encountered in equilibrium systems when truncating the hierarchy of correlations to the same order. It survives in the infinite-dimensional nonequilibrium steady state due to the amplitude of collision forces remaining $1/\sqrt{d}$ weaker than those of the self-propulsion ones, and because the flipping term ${\mathscr R}g^{(3)}$ is negligible. We now want to evaluate the last two terms in Eq.~\eqref{eq:hierarchy_second}, which in the hard-sphere limit first requires to regularize the product $g^{(3)}\bnabla_{\br'}V(\br')$. In the same spirit as in Eq.~\eqref{eq:regul-2} we can take the hard sphere limit for $V(\br')$ (for now $V(\br)$ and $V(\br'-\br)$ are kept short-ranged and regular) and we find, using Eq.~\eqref{eq:hierarchy_third}, that
\begin{equation}
\label{eq:substitution_3_body}
\begin{split}
& \lim_{\text{hard sphere}}\!\!\! 2 \int_\sigma^{+\infty}\!\!\!\! \!\!\! \dd r' g^{(3)}\partial_{r'} V(r') = \\ & \bigg(\!\!v_0(\bu' - \bu_1) \!\!- \!\!\bnabla_{\br}V(\br)\!\!-\!\!\bnabla_{\br'}V(\br'-\br)\!\!\bigg).\hat{\br}'\lim_{x \rightarrow 0^+} \int_{\sigma}^{(1+x)\sigma}  \!\!\!\!\!\!\!\!\!\!\!\!\dd r' g^{(3)}
\end{split}
\end{equation}
which holds irrespective of the $d\gg 1$ limit. We will now substitute our result for $g^{(3)}$ in terms of $g_0$ into Eq.~\eqref{eq:hierarchy_second} using first Eq.~\eqref{eq:substitution_3_body}. From the purely geometrical argument of \cite{PhysRevLett.54.2061}, we know that configurations such as those shown in Fig.~\ref{fig:tree-unlike} are exponentially rare as $d\to\infty$. If $r - \sigma = O(\sigma/d)$, which is the domain of interest of Eq.~\eqref{eq:hierarchy_second}, and given that $r'=\sigma$, we know that $\frac{||\br-\br'||}{\sigma}-1=O(1)$ except in an exponentially small fraction of the volume over which $\br'$ is integrated. It is thus safe to set $\bnabla_{\br}V(\br-\br') = \bo$ and $g_0(\br,\br';\bu_2,\bu')=1$ in Eq.~\eqref{eq:hierarchy_second}. This leads to
\begin{equation}
\begin{split}
\label{eq:mean_force}
& \rho \left\{ \int \frac{\dd \bu'}{\Omega_d}\dd \br' \, g^{(3)}(\bo,\br,\br'; \bu_1,\bu_2,\bu')  \bnabla_{\br'}V(\br') \right\} \\ & = -\frac{\rho V_d(\sigma)}{4d}g_0(\bo,\br;\bu_1,\bu_2)\left( v_0 \bu_1 + \bnabla_{\br}V(\br)\right)\left(1 + O(d^{-1/2})\right) 
\end{split}
\end{equation}
which in turn enforces $g^{(2)} = g_0$ up to $O(d^{-1/2})$ corrections as claimed in our introduction. This analytically supports the relevance of the Baxter model~\cite{baxter-1968} as a proxy for analyzing of the structure of active fluids as suggested in~\cite{Ginot-2015}. In addition, as shown in \cite{SM}, the pair product structure extends to $n$-point distributions: 
\begin{equation}
g^{(n)}(\br_1,..,\br_n ; \bu_1,..,\bu_n) = \prod_{i<j}g_0(\br_i,\br_j;\bu_i,\bu_j)
\end{equation} 
up to $O(d^{-1/2})$ corrections.
We are now in a position to determine the effective self-propulsion velocity of a tagged particle as introduced in \cite{bialke2013microscopic}. From the equation of motion~\eqref{eq:eqmo} averaged at given $\bu_i$, we define $v(\rho)$ with $\frac{\dd\langle \br_i\rangle_{i}}{\dd t }=v(\rho)\bu_i$, so that 
\begin{equation}\label{eq:vderho}
v(\rho)=v_0-\frac{\rho}{\Omega_d}\int_{\bu_j,\br_j}\!\!\!\!g^{(2)}(\br_i,\br_j;\bu_i,\bu_j)\bnabla_{\br_i}V(\br_i-\br_j)\cdot \bu_i 
\end{equation}
Using our result for $g^{(2)}$, Eq.~\eqref{eq:two_point} and Eq.~\eqref{eq:regul-2}, we arrive at a central physical result of this letter:
\begin{equation}\label{eq:qs}
v(\rho)=v_0\left(1-\frac{\rho }{\rho_\text{cr}}\right),\;\rho_{\text {cr}}=\frac{4d}{V_d(\sigma)}
\end{equation}
This immediately defines the range of validity of our calculation, such that $\rho<\rho_\text{cr}$. Indeed $\rho>\rho_\text{cr}$ would lead to a negative $v(\rho)$, which is unphysical, so that for $\rho>\rho_\text{cr}$ the system cannot be, at a microscopic level, in a homogeneous state, which echoes the findings of \cite{klamser2018thermodynamic, leticia-2018} in two-dimensional systems.  The crowding density $\rho_\text{cr}$ which controls this transition is a density scale independent of the dynamical parameters $v_0$ and $\tau$. In the analysis of existing numerical simulations, a linear function $v(\rho)$ has appeared to be an excellent fit both in two and three dimensions~\cite{PhysRevLett.108.235702,PhysRevLett.114.198301,stenhammar2014phase}. Numerics also show the vanishing of $v(\rho)$ beyond a threshold that was observed to be independent of dynamical parameters~\cite{PhysRevLett.114.198301}. We conjecture that this arrest density is the crowding density $\rho_\text{cr}$ of our calculation. Our large-dimensional prediction is that the transition occurs at a volume fraction $\phi=\rho V_d(\sigma/2)=4d2^{-d}$ which is smaller than the corresponding jamming density of hyperspheres (which goes as $6.26 d 2^{-d}$~\cite{RevModPhys.82.789} for $d\gg 1$). Paradoxically, even though the crowding threshold depends on geometry only, it is tempting to view it as a new, intrinsically dynamical, jamming scale. Finally, considering the relative motion of two particles, the quantity $v(\rho)$ not only describes their effective self-propulsion velocity, but it surprisingly also controls their effective mobility by reducing the amplitude of their direct interaction. Indeed, at given $i$ and $j$ self-propulsion velocities and positions, 
\begin{equation}
\frac{\dd}{\dd t}\langle \br_i-\br_j\rangle_{ij}=v(\rho)(\bu_i-\bu_j)-2\frac{v(\rho)}{v_0}\bnabla_{\br_i} V(\br_i-\br_j)
\end{equation}
after making use of Eq.~\eqref{eq:mean_force}.
\\
Another interesting property of active particles interacting with pairwise forces is the existence of an equation of state for the pressure $P$, in the sense that it only depends on bulk properties of the fluid. Following \cite{PhysRevLett.114.198301} the pressure in a homogeneous state is given by
\begin{equation}\label{eq:eos}
P=\rho \frac{v_0^2 \tau}{d}\frac{v(\rho)}{v_0}-\frac{\rho^2}{2d \Omega_d^2}\int_{\br,\bu_1,\bu_2}\!\!\!g^{(2)}(\bo,\br;\bu_1,\bu_2)\br\cdot\p_\br V(\br)
\end{equation}
When $\rho<\rho_\text{cr}$,  we have
\begin{equation}
\frac{P}{\sigma\rho_\text{cr}\hat{v}_0}= \frac{\hat{v}_0\tau}{\sigma} \frac{\rho}{\rho_\text{cr}}\left(1-\frac{\rho}{\rho_\text{cr}}\right)+\frac{1}{\sqrt{\pi}}\frac{\rho^2}{\rho_\text{cr}^2}
\end{equation}
This exact equation of state is consistent with numerical observations~\cite{PhysRevLett.114.198301}. It allows for spinodal instability when $\rho<\rho_\text{cr}$ and $\frac{\dd P}{\dd\rho}<0$, hence for $\hat{v}_0\tau>2\sigma/\sqrt{\pi}$ (in line with the numerical observation~\cite{stenhammar2014phase} that the instability threshold for the run length increases with dimension). When this criterion is fulfilled the spinodal region is defined by
\begin{equation}
1>\frac{\rho}{\rho_\text{cr}}>\frac{1}{2}\frac{\sqrt{\pi}\frac{\hat{v}_0\tau}{\sigma}}{ \sqrt{\pi}\frac{\hat{v}_0\tau}{\sigma}-1}
\end{equation}
The important results of this letter are threefold. i) There exists an infinite-dimensional limit in which the stationary properties of self-propelled particles interacting via a pairwise potential can be solved exactly. In the hard-sphere limit, the pair distribution function is shown to pick up a strongly attractive term at contact (in the form of a $\delta$ contribution). ii) The effective self-propulsion velocity dressed  by the interactions with other particles vanishes at a crowding density slightly smaller than the jamming one. Neither the pair distribution function nor the crowding density depend on the bare self-propulsion velocity nor on the time scale governing the decay of self-propulsion correlations. iii) These findings allow us to obtain the equation of state for self-propelled hard-spheres in the homogeneous phase, and to find the location of the spinodal preempting MIPS. The range of directions our work opens up is manifold.  To begin with physical questions of current interest, one of them stands out as a rather natural, albeit nontrivial, application of our method: hard-spheres in contact with a hard-wall are characterized by a fluid-solid surface tension the determination of which involves not only the pair distribution function~\cite{Bellemans-1962,ruben} (which we have shown how to approach), but also the density profile in the vicinity of the wall (in the spirit of \cite{ezhilan_alonso-matilla_saintillan_2015}). On a different note,  it is well-known that in equilibrium the details of the dynamics bear no influence on the stationary properties; this is of course not so out of equilibrium. Here we have studied the simplest instance of self-propelled dynamics, but hydrodynamic interactions could be incorporated {\it e.g.} in the form of an Oseen motility tensor (see \cite{charbonneau-2013} for a $d$-dimensional version). Among other extensions of interest we would like to mention, in the spirit of \cite{frisch-percus-rods-1989}, the study of self-propelled rods in which alignement interactions will now introduce an additional physical ingredient. We sense, however, that equally interesting, though more involved, research directions lie in exploring the vicinity of the crowding density (at, and beyond~\cite{leticia-2018,klamser2018thermodynamic}) and in capturing dynamical evolution~\cite{Agoritsas_2019}, allowing us to access slow dynamics properties~\cite{ni2013pushing, berthier-2019}.\\

We acknowledge very insightful exchanges with L. Berthier, M. E. Cates, D. Limmer, K. Mandadapu and J. Tailleur.


\bibliography{biblio-activemf}


\end{document}


\title{\bf{Active hard-spheres in infinitely many dimensions: Supplemental Material}}
\author{Thibaut Arnoulx de Pirey}
\affiliation{Université de Paris, Laboratoire Mati\`ere et Syst\`emes Complexes (MSC), UMR 7057 CNRS, F-75205 Paris, France}

\author{Gustavo Lozano}
\affiliation{Departmento de F\'isica, Facultad de Ciencias Exactas y Naturales,
Universidad de Buenos Aires, Ciudad Universitaria, Pabell\'on I, 1428 Buenos Aires, Argentina
}

\author{Fr\'ed\'eric van Wijland}
\affiliation{Université de Paris, Laboratoire Mati\`ere et Syst\`emes Complexes (MSC), UMR 7057 CNRS, F-75205 Paris, France}

\maketitle


\section{The infinite dimensional limit in and out of equilibrium}

The diffusion constant of a noninteracting particle is $\frac{v_0^2\tau}{d}$ and we keep it fixed. The exact results we have presented in the main text hold in the highly ballistic limit where $\tau$ is fixed and $v_0 = \sqrt{d}\hat{v}_0$, with $\hat{v}_0$ fixed. Other scalings are consistent with a nonequilibrium limiting process. For instance, still at fixed diffusion constant, we could have chosen $v_0 = d^\alpha v_0'$ and $\tau = d^{1-2\alpha}\tau'$, with $v_0'$ and $\tau'$ fixed, and $\alpha$ arbitrary. In this case, the run length is $\ell= d^{1-\alpha} v_0'\tau'$. The other interesting length scale associated with the hard-sphere interaction is the typical skidding length $\ell_s$ during a two-body collision (defined as the length run by a particle skidding around another in the absence of any tumble). Since grazing collisions dominate the $d \gg 1$ limit, this skidding distance (see Fig. 1 of the main text) typically scales as $\ell_s = \sigma/\sqrt{d}$. At this stage, we may want to compare $\ell$ with respect to $\ell_s$ . Equilibrium is recovered when $\ell\ll\ell_s$ , which enforces $\alpha> 3/2$. Right at $\alpha=3/2$, tumbling and collisions play equally important roles in the equations of motion. While this might seem an appealing scaling to work with, it turns out that it suppresses the possibility of MIPS understood as the spontaneous destabilization of the homogeneous phase. By contrast, our choice of $\alpha=1/2$ is the only one that results in an equation of state where both the hard-core repulsion and the effective attraction play equally important roles. Note that once the highly ballistic limit with $\alpha=1/2$ is chosen, it is not possible to recover the equilibrium physics of hard-spheres by taking the $v_0'\tau\to 0$ limit afterwards.

\section{Evolution of the orientational degrees of freedom}
In the main text we have asserted that the specifics of the active dynamics imparted on the particles did not alter the validity of our results. In this section we establish the connections, in infinitely many dimensions, between run and tumble particles (RTP), active Brownian (ABP) and active Ornestein-Uhlenbeck (AOUP) ones. The purpose of this section is to gather the relevant information from existing literature~\cite{cates2015motility,PhysRevE.90.012111} in order to properly define the infinite dimensional limit of alternative models frequently encountered in the description of active particles. For RTPs, the active force acting on a particle appears through a contribution $v_0\bu$, where $\bu$ is a unit vector uniformly picking a random  orientation at random times drawn from a Poisson distribution with density $\tau^{-1}$. The probability $p(\bu, t)$ that a particle has orientation $\bu$ evolves according to
\begin{equation}\label{eq:eqRTP}
\p_tp(\bu,t)={\mathscr R}p=\frac{1}{\tau}\int_{\bu'}\frac{p(\bu',t)}{\Omega_d}-\frac{1}{\tau}p(\bu,t)
\end{equation}
and in the stationary state $p_\text{st}(\bu)=\frac{1}{\Omega_d}$ is uniform (here $\Omega_d=\frac{2 \pi^{d/2}}{\Gamma(d/2)}$ refers to the solid angle in $d$ dimensions). Our Eq.~\eqref{eq:eqRTP} defines the operator $\mathscr R$ that appears in the main text, and it allows us to show that $\langle\bu(t)\cdot\bu(t')\rangle=\ee^{-|t-t'|/\tau}$.  Had we chosen to work with active Brownian particles (ABPs) instead, the vector $\bu$ would have evolved according to
\begin{equation}
\frac{\dd \bu}{\dd t}\displaystyle \underset{\text{Stratonovich}}{=}-\bu(\bu\cdot\beeta)+\beeta\underset{\text{Ito}}{=}-(d-1)D_r\bu-\bu(\bu\cdot\beeta)+\beeta
\end{equation}
where the components $\eta^\alpha$ of the Gaussian white noise have correlations $\langle\eta^\alpha (t)\eta^\beta(t')\rangle=2D_r \delta^{\alpha\beta}\delta(t-t')$. This shows that $\langle\bu(t)\cdot\bu(t')\rangle=\ee^{-(d-1)D_r|t-t'|}$ and that
\begin{equation}
\p_tp(\bu, t) = {\mathscr R}p = D_r \Delta_\bu p
\end{equation}
where the Laplacian $\Delta_\bu$ is on the unit sphere. Hence our persistence time $\tau$ must be identified with $\tau=\frac{1}{(d-1)D_r}$ and for ABPs our large dimensionality analysis would require to work at $D_r \to 0$ while preserving $(d-1)D_r$ finite. The stationary distribution $p_\text{st}(\bu)$ is identical to that of RTPs. Finally, for active Ornstein-Uhlenbeck particles (AOUPs), which are most convenient in numerical simulations (see {\it e.g.}~\cite{flenner2016nonequilibrium}), the vector $\bu$ does not have a fixed norm. It evolves according to
\begin{equation}
\frac{\dd\bu}{\dd t}=-\frac{\bu}{\tau}+\sqrt{\frac{2}{d\tau}}\bxi
\end{equation}
where now the Gaussian white noise has correlations $\langle\xi^\alpha(t)\xi^\beta(t')\rangle=\delta^{\alpha\beta}\delta(t-t')$. This defines a standard Ornstein-Uhlenbeck process, and $p(\bu, t)$ then evolves according to
\begin{equation}
\p_t p={\mathscr R}p=\frac{1}{\tau}\left(\p_\bu\cdot(\bu p)+\frac{1}{d}\p_\bu^2 p\right)
\end{equation}
from which one recovers $\langle\bu(t)\cdot\bu(t')\rangle=\ee^{-|t-t'|/\tau}$. The stationary distribution $p_\text{st}(\bu)=\frac{1}{\sqrt{2\pi/d}}\ee^{-d\bu^2/2}$, once integrated over the proper volume element $\int_\bu p_\text{st}(\bu)\dots=\int\dd\Omega_\bu\dd u \,u^{d-1}\frac{1}{\sqrt{2\pi/d}}\ee^{-d\bu^2/2}\ldots$ shows that the value of $\bu$ that eventually dominate statistics are such that $\ln u - u^2/2$ is the largest, that is such that $||\bu|| = 1$ , up to vanishingly small fluctuations as $d\to\infty$, making a link with RTPs and ABPs.

In all these cases, flips will lead to negligible contributions in equations for correlation functions as the dimension goes to infinity. The conclusions presented in the main text for RTPs thus extend to ABPs and AOUPs.

\section{Solving the two-body problem}
\begin{figure}

\begin{overpic}[totalheight=5cm]{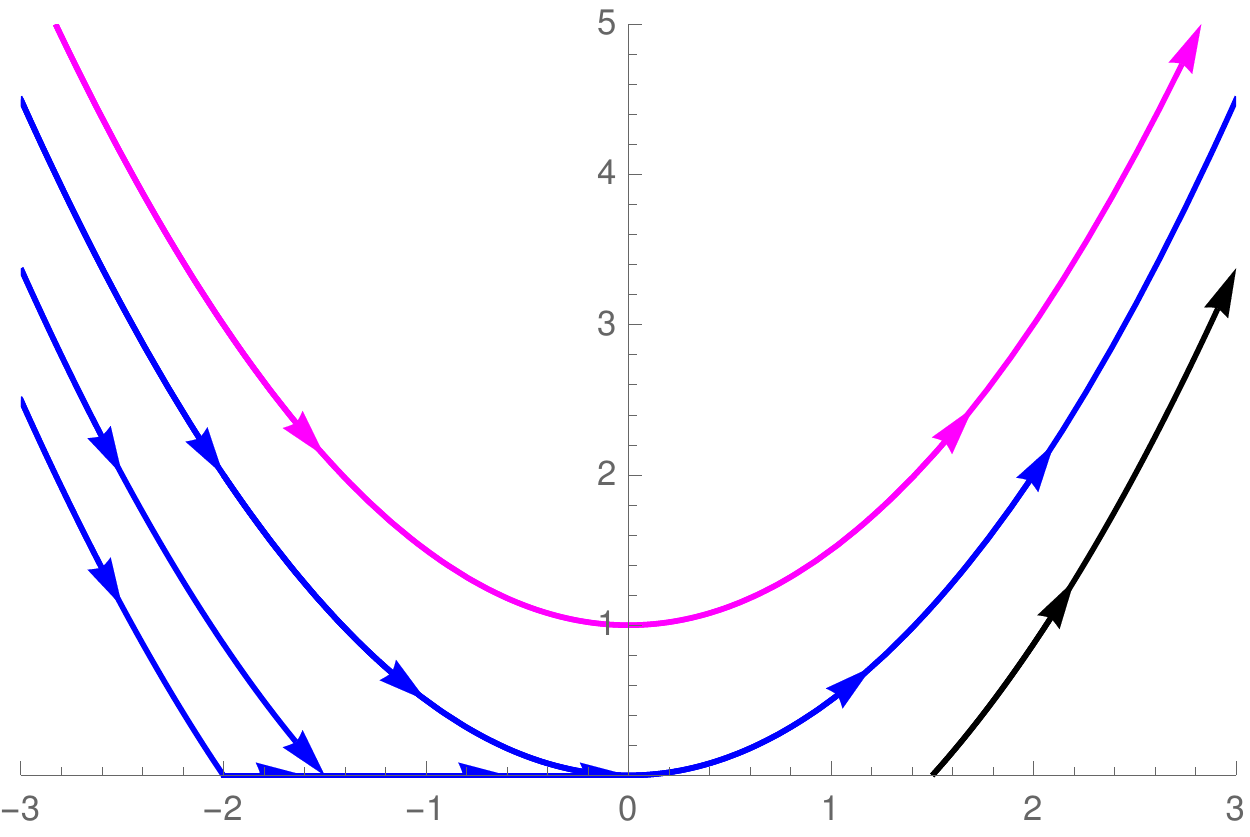}

 \put (40,45) {\textcolor{magenta}{Region 2}}
 \put (83,12) {Region 3}
 \put (0,10) {\textcolor{blue}{Region 1}}
 \put (49,67){$z$}
 \put (100,4) {$x$}
\end{overpic}
  \caption{Sample trajectories of the relative particle of a two particle system in the $(x,z)$ plane are shown. These are also the characteristics of the partial differential equation Eq.~\eqref{eq:regular} we wish to solve. Their equation $z= p + x^2/2$ is indexed by $p$ defined from the impact parameter $b$ by $b= \sigma(1 + p/d)$. For $p<0$, on the lhs of the vertical axis (region 1, blue lines), they show three trajectories leading to a collision. The two particles then skid around each other until $x$ reaches $0$. For $p > 0$ (region 2, magenta line) they describe a trajectory without a collision.  The outgoing characteristics (region 3, black) would correspond to particles leaving the collision with $x > 0$, which never occurs when flips are neglected. \label{fig:karact}}
\end{figure}

In the main text, we defined $g_0(\bo,\br;\bu_1,\bu_2)$ the stationary two-point function of the two-body problem for finding one particle at $\bo$ with orientation $\bu_1$ and one particle at $\br$ with orientation $\bu_2$. It is a solution of 
\begin{equation}
\label{eq:two_body}
-v_0 (\bu_2 - \bu_1) \cdot \bnabla_{\br}g_0 + 2 \bnabla_{\br}\cdot\left(g_0 \bnabla_{\br}V(\br)\right) +  {\mathscr R}g_0 = 0
\end{equation}
For an arbitrary repulsive potential, the two particles cannot get closer than $r\equiv||\br|| = r^*$ where $r^*$ is given by $\left. \partial_r V \right|_{r^*} = - 2 v_0$. This corresponds indeed to the case of an exactly head-on collision with $\bu_1 = -\bu_2 = \hat{\br}$. During the underlying dynamical process, only an exactly head-on collision can bring $r$ down to $r^*$. Thus $g_0(\bo,\br;\bu_1,\bu_2)$ must satisfy the following two boundary conditions:
\begin{equation}
\left\{
\begin{split}
g_0(\bo,\br;\bu_1,\bu_2) & = 0 \text{ for } r=r^* \\
g_0(\bo,\br;\bu_1,\bu_2) & = 1 \text{ as } r \to  \infty
\end{split}
\right.
\end{equation}
We are interested in the hard-sphere limit of Eq.~\eqref{eq:two_body} which we rewrite as
\begin{equation}\label{eq:flux}
-v_0 \left[(\bu_2 - \bu_1) \cdot \hat{\br} \right]\partial_{r}g_0 - \frac{v_0}{r} (\bu_2 - \bu_1) \cdot\bnabla_{\hat{\br}}g_0 + \frac{2}{r^{d-1}} \partial_{r}\left(r^{d-1}g_0 \partial_{r}V(r)\right) +  {\mathscr R}g_0 = 0
\end{equation}
Then we multiply Eq.~\eqref{eq:flux} by $r^{d-1}$ and integrate over $r^* < r < r^*(1+\epsilon)$ where $\epsilon>0$ is arbitrary. This yields,
\begin{equation}
\label{eq:integrate}
\begin{split}
& -v_0\left[\left(\bu_2-\bu_1\right)\cdot \hat{\br}\right]\left(r^*(1+\epsilon)\right)^{d-1}g_0(\bo,\hat{\br},r^*(1+\epsilon))+(d-1)v_0\left[\left(\bu_2-\bu_1\right)\cdot \hat{\br}\right]\int_{r^*}^{r^*(1+\epsilon)}\dd r \, r^{d-2}g_0 \\
& -v_0 \left(\bu_2-\bu_1\right)\cdot \bnabla_{\hat{\br}}\int_{r^*}^{r^*(1+\epsilon)}\dd r \, r^{d-2}g_0 + 2 \left(r^*(1+\epsilon)\right)^{d-1}g_0(\bo,\hat{\br},r^*(1+\epsilon)) \partial_r V(r^*(1+\epsilon)) \\ & + {\mathscr R}\int_{r^*}^{r^*(1+\epsilon)}\dd r \, r^{d-1}g_0 = 0
\end{split}
\end{equation}
We are now in a position to study the hard-sphere limit, the first consequence of which is to send $r^* \to \sigma$, the diameter of a particle. Moreover, at fixed $\epsilon$, $\partial_r V(r^*(1+\epsilon))$ goes to $0$. After the hard-sphere limit we take the $\epsilon \to 0$ limit in Eq.~\eqref{eq:integrate} which leads to
\begin{equation}
\begin{split}
\label{eq:HS_limit}
\lim_{\epsilon \to 0}& \left[-v_0\left[\left(\bu_2-\bu_1\right)\cdot \hat{\br}\right]g_0(\bo,\hat{\br},\sigma(1+\epsilon))+\frac{(d-1)v_0}{\sigma}\left[\left(\bu_2-\bu_1\right)\cdot \hat{\br}\right]\int_{\sigma}^{\sigma(1+\epsilon)}\dd r \, g_0 
\right. \\ & \left. -\frac{v_0}{\sigma} \left(\bu_2-\bu_1\right)\cdot \bnabla_{\hat{\br}}\int_{\sigma}^{\sigma(1+\epsilon)}\dd r \, g_0 + {\mathscr R}\int_{\sigma}^{\sigma(1+\epsilon)}\dd r \, g_0\right] = 0
\end{split}
\end{equation}
This proves that the solution for $g_0$ displays a delta peak accumulation at contact with $g_0(\bo,\br;\bu_1,\bu_2) = \Theta(r-\sigma)f(\br;\bu_1,\bu_2) + \delta\left(\frac{r-\sigma}{\sigma}\right)\Gamma(\hat{\br};\bu_1,\bu_2)$ and that the functions $f$ and $\Gamma$ satisfy 
\begin{equation}
\label{eq:regular}
-v_0 \left(\bu_2 - \bu_1\right)\cdot\bnabla_\br f + ({\mathscr R}f) = 0
\end{equation}
and
\begin{equation}\label{eq:singular}
-v_0(\bu_2-\bu_1)\cdot\left[f(\sigma \hat{\br};\bu_1,\bu_2)\hat{\br} +\nabla_{\hat{\br}}\Gamma-(d-1)\Gamma\hat{\br}\right] + ({\mathscr R}\Gamma) = 0
\end{equation}
as claimed in the main text in Eqs.~(4) and (5). The quantity $\Gamma$ is non-vanishing only for colliding particles  with $(\bu_2-\bu_1)\cdot\br<0$. Before solving these equations in the large dimensional limit, we use Eq.~\eqref{eq:integrate} to prove that the product $g_0(\bo,\br;\bu_1,\bu_2)\bnabla_{\br}V(\br)$ is well defined in the hard-sphere limit. In equilibrium, one could prove this by noting that $g_0(\bo,\br;\bu_1,\bu_2)\bnabla_{\br}V(\br) \propto \bnabla_{\br}\left(\ee^{-\beta V(\br)}\right)$ which is ambiguity-free. In order to endow $ g_0\bnabla V$ with a definite mathematical meaning, we integrate Eq.~\eqref{eq:integrate} over $0 < \epsilon < \epsilon'$ with $\epsilon'>0$ fixed. We first take the hard-sphere limit and then the $\epsilon' \to 0$ limit. The $2^{nd}$, $3^{rd}$ and $5^{th}$ term of Eq.~\eqref{eq:integrate} yield vanishing contributions and the remaining terms give 
\begin{equation}\label{eq:regul-2}
\begin{split}
\lim_{\text{hard-sphere}} \int_\sigma^{+\infty} \dd r g_0(\bo,\br;\bu_1,\bu_2)\partial_r V(r) = \frac{v_0}{2}\left[\left(\bu_2-\bu_1\right).\hat{\br}\right] \Gamma(\hat{\br};\bu_1,\bu_2)
\end{split}
\end{equation}
as also claimed in the main text in Eq.~(8). More generally, in terms of distributions, we have that
\begin{equation}\label{eq:regul-distrib}
\begin{split}
\lim_{\text{hard-sphere}} g_0(\bo,\br;\bu_1,\bu_2)\partial_r V(r) = \frac{v_0}{2}\left[\left(\bu_2-\bu_1\right).\hat{\br}\right] \Gamma(\hat{\br};\bu_1,\bu_2)\delta(r-\sigma)
\end{split}
\end{equation}
which is, for instance, useful in determining the equation of state Eq.~(17) of the main text. In order to compute $f$ and $\Gamma$, we now solve the coupled equations Eqs.~\eqref{eq:regular} and \eqref{eq:singular}. In the $d \gg 1$ limit, ${\mathscr R}f$ scales as $O(1)$. On the contrary, $v_0 \left(\bu_2 - \bu_1\right)\cdot\bnabla_\br f$ scales as $O(d)$. Indeed, $v_0 \sim O(\sqrt{d})$, $||\bnabla_\br f|| \sim O(d)$ because spatial variations occur on scales of order $O(\sigma/d)$ and there is a $1/\sqrt{d}$ factor coming from the typical value of the dot product between two unit vectors. This ratio of $O(d)$ between the streaming and the flipping terms stems from the highly ballistic regime adopted in our definition of the large $d$ limit (as discussed in the introduction of the main text). Indeed the typical skidding distance is $O(\sigma/\sqrt{d})$, while the run length is $\sqrt{d}\hat{v}_0\tau$. In both Eq.~\eqref{eq:regular} and Eq.~\eqref{eq:singular}, flipping terms can be omitted. We now define $\theta$ the angle between $\left(\bu_1 - \bu_2\right)$ and $\br$, $x$ such that $x = \sqrt{d}\cos(\theta)$ and $z = \frac{d(r-\sigma)}{\sigma}$. We also introduce $\hat{\Gamma}(x) = d \Gamma(x)$. In these rescaled coordinates we have
\begin{equation}\label{eq:free}
\left\{
\begin{split}
& x\partial_z f(z,x) + \partial_x f(z,x) = 0 \\
& -x f(0,x) + x \hat{\Gamma}(x) - \partial_x \hat{\Gamma}(x) = 0
\end{split}
\right.
\end{equation}
with $\hat{\Gamma}(x)=0$ for $x>0$. The first equation of Eq.~\eqref{eq:free} teaches us that $f(z,x)$ is constant along the lines $z - \frac{1}{2}x^2 = \text{cst}$ which correspond to free streaming trajectories, as depicted in Fig.~\ref{fig:karact}. Thus $f(z,x) = 1$ for $z - x^2/2 > 0$ (trajectories with no collision) and for $z - x^2/2 < 0$ with $x <0$ (trajectories leading to a collision). On the other hand, for $x > 0$, $f(0,x) = 0$ which implies that $f(z,x) = 0$ for $z - x^2/2 < 0$ with $x >0$ (trajectories of particles leaving the collision). For $d$ large but finite,  $f(z,x)$ actually scales as $1/d$ in that region of phase space where $z - x^2/2 < 0$, $x >0$. In addition, we find $\hat{\Gamma}(x) = \sigma$ for $x <0$. We thus recover the result Eq.~(6) of the main text, namely
\begin{equation}
\begin{split}
\label{eq:two_point}
g_0(\bo,\br;\bu_1,\bu_2) = \, & \Theta(r - \sigma)\left[1 - \Theta(\cos\theta)\Theta(\sigma - r \sin\theta)\right] + \Theta(-\cos\theta)\delta\left(\frac{d(r-\sigma)}{\sigma}\right)
\end{split}
\end{equation}

\section{Loop configurations and the infinite dimensional limit}
\begin{figure}[h]
\begin{overpic}[totalheight=3cm]{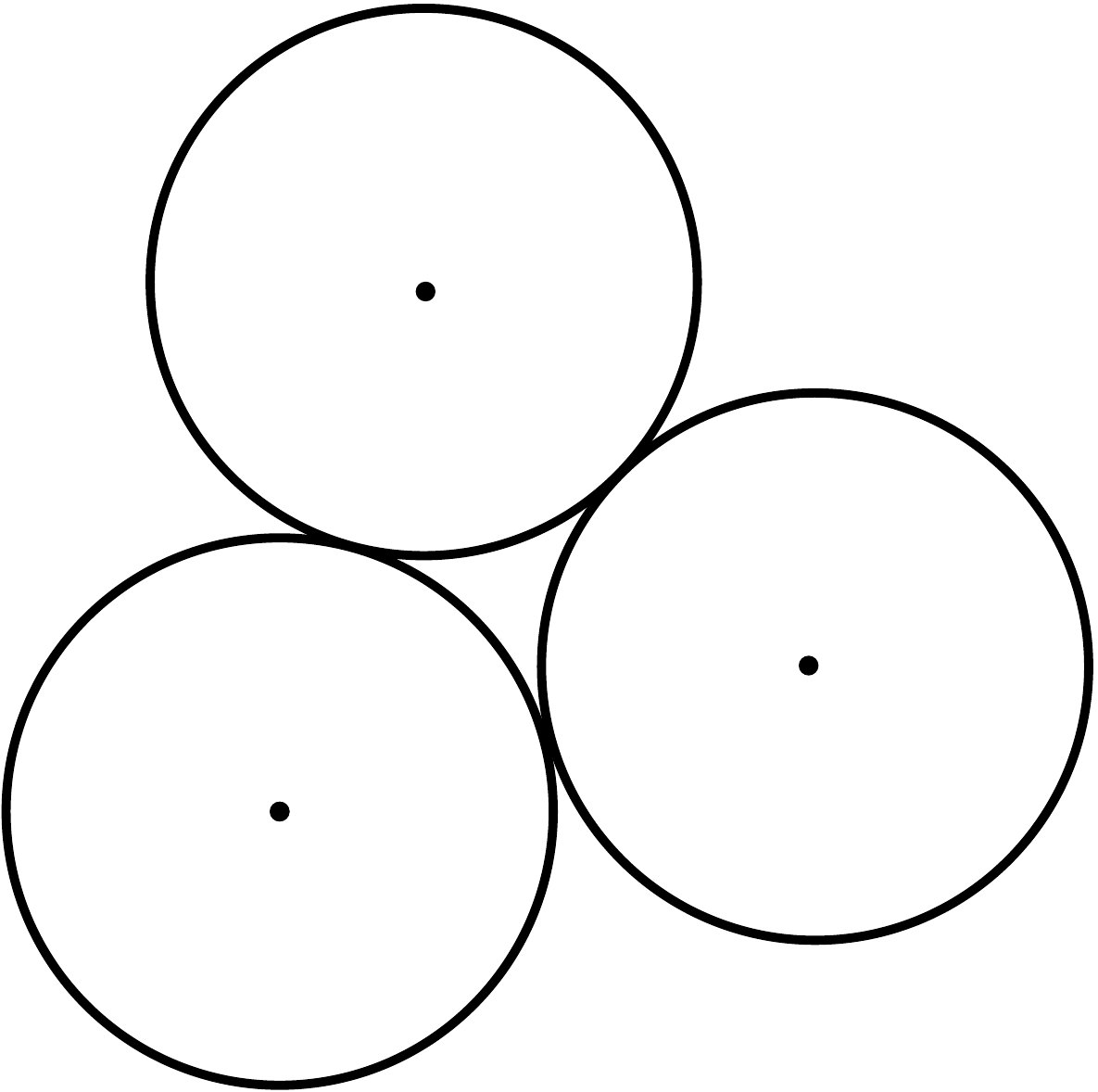}
\put (15,22) {$\bo$}
\put (75,30) {$\br$}
\put (40,75) {$\br'$}
\end{overpic}
\caption{Three hard-spheres at positions $\bo$, $\br$, $\br'$ forming a loop of contacts, thus being in an unlikely  spatial configuration as $d\to\infty$. \label{fig:tree-unlike-SM}}
\end{figure}

Here we want to argue why three-sphere configurations such as shown in Fig.~2 of the main text, or Fig.~\ref{fig:tree-unlike-SM} of the present Supplemental Material, contribute an exponentially small correction (in $d$) to our results.
We assume that we have 2 spheres of diameter $\sigma$ at positions $\bo$ and $\br$ in contact with each other and that a third one is in contact with at least one of them. The set of available positions of the third particle defines a surface  with area $\Omega_d \sigma^{d-1} < \mathcal{S} < 2 \Omega_d \sigma^{d-1}$. We are now interested in $s$ the fraction of the total area $\mathcal S$ occupied by configurations in which the third particle is in contact with both. We have:
\begin{equation}
\begin{split}
s & = \frac{1}{\mathcal{S}}\int \dd^d \br' \delta(r' - \sigma) \delta\left(\frac{||\br' - \br|| - \sigma}{\sigma}\right) \\
& = \frac{\Omega_{d-1}\sigma^{d-1}}{\mathcal{S}}\int_0^\pi \dd \theta \sin^{d-2}\theta \, \, \delta\left(\sqrt{2}\sqrt{1-\cos^2\theta}-1\right) \\
& = \frac{\Omega_{d-1}\sigma^{d-1}}{\sqrt{2}\mathcal{S}}\left(\frac{1}{2}\right)^{\frac{d-3}{2}}
\end{split}
\end{equation}
so that the fraction of the total area occupied by loop configurations  is exponentially small in $d$ as $d \to \infty$. These conclusions  extend to the case of nearby particles--in the sense that their relative separation $r$ is $\sigma < r < \sigma(1+z/d)$ where $z$ is some positive $O(1)$ constant--or to more than 3 particles. Since loop configurations only occupy an exponentially small volume fraction of space, they can be neglected in the limit $d \to \infty$.
\section{Solving the hierarchy of correlations in the infinite-dimensional limit}
In a nonequilibrium steady-state, there is of course no known general form for the distribution. Our goal in this section is to show that it can actually be found in the $d\gg 1$ limit for self-propelled hard-spheres. We want to establish Eq.~(13) of the main text in which the explicit form of the stationary distribution is stated.

In the thermodynamic limit, our starting point is the infinite hierarchy of equations relating $n$-body correlation functions $g^{(n)}(\br_1,\dots,\br_n;\bu_1,\dots,\bu_n)$ to $(n+1)$-body ones, known as the Bogoliubov–Born–Green–Kirkwood–Yvon (BBGKY) hierarchy. For our model it reads
\begin{equation}
\begin{split}
& -v_0 \sum_{i=1}^n \bu_i.\bnabla_{\br_i}g^{(n)} + \sum_{i=1}^n \sum_{j  \neq i} \bnabla_{\br_i}\left(g^{(n)} \bnabla_{\br_i}V(\br_i - \br_j)\right) + \frac{1}{\tau} \sum_{i=1}^n \left(\int \frac{\dd \bu_i'}{\Omega_d} g^{(n)} - g^{(n)} \right) \\ & + \rho \sum_{i=1}^n \bnabla_{\br_i} \int \dd \br' \, \frac{\dd\bu'}{\Omega_d} g^{(n+1)}(\br_1,...,\br_n, \br' ; \bu_1,...,\bu_n, \bu') \bnabla_{\br_i}V(\br_i - \br')  = 0
\end{split}
\end{equation}
We find it convenient to introduce an infinite, virial-like, series expansion of the correlation functions in powers of the combination $\frac{\rho V_d(\sigma)}{d}$, which is $O(1)$ in the regimes studied in this paper. For the two-point function, we thus write
\begin{equation}\label{eq:2point-virial}
g^{(2)}(\br_1,\br_2;\bu_1,\bu_2) = \sum_{p=0}^{+ \infty} \left(\frac{\rho V_d(\sigma)}{d}\right)^p g^{(2)}_p(\br_1,\br_2;\bu_1,\bu_2) 
\end{equation}
We later self-consistently show that $\frac{\rho V_d(\sigma)}{d}$ is the correct expansion parameter so that for all $p \geq 0$ the functions $g^{(2)}_p(\br_1,\br_2;\bu_1,\bu_2) $ remain $O(1)$ (or less) as $d$ increases. By truncating the hierarchy to order $q \geq 2$, we have access to all $g^{(2)}_p(\br_1,\br_2;\bu_1,\bu_2)$ for $p \leq q - 2$. We thus start by assuming such a truncation holds, \textit{i.e.} we look for $g^{(q)}(\br_1,\dots,\br_q;\bu_1,\dots,\bu_q)$ such that 
\begin{equation}
\label{eq:BBGKY}
-v_0 \sum_{i} \bu_i \cdot \bnabla_{\br_i} g^{(q)} + \sum_{i\neq j} \bnabla_{\br_i}\cdot\left( g^{(q)}\bnabla_{\br_i}V(\br_i-\br_j)\right) + \frac{1}{\tau}\sum_{i} \left( \int \frac{\dd \bu_i'}{\Omega_d}g^{(q)} - g^{(q)}\right) = 0
\end{equation}
We look for a solution of the previous equation in the form 
\begin{equation}
\label{eq:ansatz}
g^{(q)}(\br_1,\dots,\br_{q};\bu_1,\dots,\bu_{q}) = \left[\prod_{i > j}g_0(\br_i,\br_j;\bu_i,\bu_j)\right]\bigg(1 + K(\br_1,\dots,\br_{q};\bu_1,\dots,\bu_{q})\bigg)
\end{equation}
where the product runs over $1 \leq i \leq q$ and $1 \leq j \leq q$.
In the following, in order to lighten notations, we will write $g_0(\br_i,\br_j;\bu_i,\bu_j)= g_0(i,j)$. By inserting Eq.~\eqref{eq:ansatz} into Eq.~\eqref{eq:BBGKY} we eventually obtain
\begin{equation}\label{eq:pasbel}
\begin{split}
& -\frac{(1+K)}{\tau}\sum_{i > j}\left(\displaystyle \prod_{\substack{p > m \\ (p,m) \neq (i,j)}}g_0(p,m)\right)\left(\int \frac{\dd \bu_i'}{\Omega_d}g_0(i,j) + \int \frac{\dd \bu_j'}{\Omega_d}g_0(i,j) - 2g_0(i,j)\right) \\ & + \frac{1}{\tau}\sum_i\left(\int \frac{\dd \bu_i'}{\Omega_d}(1+K)\left(\displaystyle \prod_{\substack{p > m}}g_0(p,m)\right) -  (1+K)\left(\prod_{\substack{p > m}}g_0(p,m)\right)\right) \\ & - v_0 \sum_i \left[\displaystyle \prod_{\substack{p > m}}g_0(p,m)\right] \bu_i \cdot \bnabla_{\br_i}K + \sum_{i \neq j} \left(\displaystyle \prod_{\substack{p > m}}g_0(p,m)\right) \bnabla_{\br_i}K \cdot \bnabla_{\br_i}V(\br_i-\br_j) \\ & + \sum_{i > j > k} \left[\left(\displaystyle \prod_{\substack{p > m \\ (p,m) \neq (i,k) }}g_0(p,m)\right)(1+K)\bnabla_{\br_i}g_0(i,k)\cdot \bnabla_{\br_i}V(\br_i-\br_j) + (i \leftrightarrow j \leftrightarrow k)\right] = 0
\end{split}
\end{equation}
Let us now determine the order in $d$ of the various terms entering Eq.~\eqref{eq:pasbel}. In the first two lines they all are $O(1)$. Both terms in the third line are of order $O(K d)$ (the order of $K$ is yet unspecified). In order to see this, we use first that $v_0=\sqrt{d}\hat{v}_0$, $||\bnabla V||=O(1)$ and $||\bnabla K||=O(d K)$. A dot product between typically independent vectors comes with a $d^{-1/2}$ amplitude, hence the overall $O(d K)$ of the first contribution in the third line of Eq.~\eqref{eq:pasbel}. The other one features a dot product between colinear vectors, which altogether leads to the same $O(d K)$ for the second contribution. By means of a similar reasoning, the terms in the last line of Eq.~\eqref{eq:pasbel} are found to be of order $O(\sqrt{d})$. Thus, the unknown function $K$ scales as $O(d^{-1/2})$. As a conclusion, in this truncated scheme, we have that
\begin{equation}
\label{eq:p_point}
g^{(q)}(\br_1,\dots,\br_{q};\bu_1,\dots,\bu_{q}) = \left[\prod_{i > j}g_0(\br_i,\br_j;\bu_i,\bu_j)\right]\bigg(1 + \frac{\hat{K}(\br_1,\dots,\br_{q};\bu_1,\dots,\bu_{q})}{\sqrt{d}}\bigg)
\end{equation}
with $\hat{K}$ an $O(1)$ function as claimed in the main text after Eq.~(10) for a truncation to the level of the third equation. We now need to evaluate the contribution of $q$-body correlations in the equation for $(q-1)$-body ones and thus we need to compute, in the hard-sphere limit, quantities such as
\begin{equation}
\label{eq:force}
\rho \bnabla_{\br_1}\cdot\int \dd \br' \frac{\dd \bu'}{\Omega_d}g^{(q)}(\br_1,\dots,\br';\bu_1,\dots,\bu')\bnabla_{\br_1}V(\br_1-\br')
\end{equation}
We follow the route shown above in Eq.~\eqref{eq:regul-distrib} for $g_0$, in order to make sense of the product $g^{(q)}\bnabla_{\br'}V(\br')$ in the limit where $V(\br')$ becomes a hard-sphere potential. We go back to Eq.~\eqref{eq:BBGKY}, where we change variables  $\bx_i = \br_i - \br_1$ for $1<i<q$ and $\br' = \br_{q}-\br_1$. We obtain
\begin{equation}
\begin{split}
& - v_0 \sum_{i=2}^{q-1} \left(\bu_i - \bu_1\right)\cdot\bnabla_{\bx_i}g^{(q)} - v_0 \left(\bu_{q}-\bu_1\right)\cdot\bnabla_{\br'} g^{(q)} + \frac{1}{\tau} \sum_{i=1}^{q} \left(\int \frac{\dd \bu_i'}{\Omega_d} g^{(q)} - g^{(q)}\right) \\ & + \sum_{i=2}^{q-1} \bnabla_{\bx_i}\cdot\left[g^{(q)}\left(\displaystyle \sum_{\substack{j = 2 \\ j \neq i}}^{q-1} \bnabla_{\bx_i}V(\bx_i - \bx_j) + \bnabla_{\bx_i}V(\bx_i - \br') + 2 \bnabla_{\bx_i}V(\bx_i) + \sum_{j \neq i}^{q-1} \bnabla_{\bx_j}V(\bx_j) + \bnabla_{\br'}V(\br') \right)\right] \\ & + \bnabla_{\br'}\cdot\left[g^{(q)}\left(\displaystyle \sum_{i = 2}^{q-1} \bnabla_{\br'}V(\br' - \bx_i) + 2 \bnabla_{\br'}V(\br') + \sum_{i = 2}^{q-1} \bnabla_{\bx_i}V(\bx_i) \right)\right] = 0
\end{split}
\end{equation}
At fixed $\bx_i$'s and $\bu_i$'s, we integrate the previous equation over $\br' \in \delta \Omega$, a small conical slab of angular aperture $\delta \Omega_d$ and for $r'$ between $\sigma(1- \epsilon)$ and $\sigma(1+\epsilon)$. We then integrate over $\epsilon$ between $0$ and $\epsilon'$ and we take the hard-sphere limit only for the  $V(\br')$ potential (keeping the other pair potentials short-ranged and regular) at fixed $\epsilon'$. Eventually, we take the limit $\epsilon' \rightarrow 0$. This yields
\begin{equation}\label{eq:regul-q}
\begin{split}
& 2 \!\!\!\!\!\!\lim_{\text{hard-sphere}} \int_\sigma^{+\infty} \dd r' g^{(q)}(\bo,\bx_2,\dots,\bx_{q-1},\hat{\br}',r';\bu_1,\dots,\bu_{q})\partial_{r'}V(r')=\\   & \lim_{\epsilon' \rightarrow 0} \int_\sigma^{\sigma(1+\epsilon')}\!\!\! \!\!\!\!\!\!\dd r'\left(v_0(\bu_{q} - \bu_1) \! - \! \sum_{i=2}^{q-1}\bnabla_{\br'}V(\br'-\bx_i) \! - \! \sum_{i=2}^{q-1}\bnabla_{\bx_i}V(\bx_i) \right)\cdot\hat{\br}' \times\ldots\\
&\;\;\;\;\;\;\;\;\;\;\;\;\ldots\times g^{(q)}(\bo,\bx_2,\dots,\bx_{q-1},\hat{\br}',r';\bu_1,\dots,\bu_{q})
\end{split}
\end{equation}
This result is valid independently of the large dimensional limit. Had we directly substituted the solution found in Eq.~\eqref{eq:p_point} into Eq.~\eqref{eq:force}, the $\bnabla_{\bx_i}V(\bx_i)$ terms would simply be missing in Eq.~\eqref{eq:regul-q}. Their norm is of order $O(1)$, which is smaller by a factor $d^{-1/2}$ than the norm of the leading term $v_0(\bu_q-\bu_1)$, hence there is no mathematical inconsistency between Eq.~\eqref{eq:regul-q} and Eq.~\eqref{eq:p_point}. Yet, a word of caution is in order. If a given vector has a norm  $O(d^{\alpha})$, its dot product with a unit vector can be either $O(d^{\alpha})$ (if they are colinear) or $O(d^{\alpha-\frac 12})$ (which is typically the case in high dimension). These $\bnabla_{\bx_i}V(\bx_i)$ terms will prove essential when Eq.~\eqref{eq:regul-q} is dotted with the $\bnabla$ operator as required in Eq.~\eqref{eq:force}.

We now use Eq.~\eqref{eq:p_point} to determine Eq.~\eqref{eq:force}. The corresponding integral is restricted over regions of space where $||\br_1-\br'||=\sigma$. Moreover, we assume that all the particles at $\br_1,\ldots,\br_{q-1}$  form one connected cluster of nearby particles. This is indeed the domain of interest of the $(q-1)$-body correlation functions. Therefore, except over an exponentially small fraction of the integration volume, we can set $V(\br_i-\br') = 0$ and $g_0(\br_i,\br';\bu_i,\bu_{q}) = 1$ for $i>1$ when evaluating the term shown in Eq.~\eqref{eq:force}. This leads, up to exponentially small corrections in $d$, to
\begin{equation}
\begin{split}
& \rho \int \dd \br' \frac{\dd \bu'}{\Omega_d}g^{(q)}(\br_1,\dots,\br';\bu_1,\dots,\bu')\bnabla_{\br_1}V(\br_1-\br') = \\ & \frac{\rho \sigma^d}{2 d}\left[\prod_{i < j} g_0(\br_i,\br_j;\bu_i,\bu_j)\right]\int \dd \hat{\br}' \frac{\dd \bu'}{\Omega_d} \hat{\br}' \left[\left(v_0(\bu' - \bu_1) + \sum_{i>1} \bnabla_{\br_1}V(\br_1-\br_i)\right)\cdot \hat{\br}'\right] \Theta(\hat{\br}'\cdot(\bu_1-\bu'))\left(1 + \frac{\hat{K}}{\sqrt{d}}\right) \\ & = - \frac{\rho V_d(\sigma)}{4 d}\left[\prod_{i < j} g_0(\br_i,\br_j;\bu_i,\bu_j)\right]\left(-v_0 \bu_1 + \sum_{i>1} \bnabla_{\br_1}V(\br_1-\br_i)\right) \\ & + \frac{\rho \sigma^d}{2 d^{3/2}}\left[\prod_{i < j} g_0(\br_i,\br_j;\bu_i,\bu_j)\right]\int \dd \hat{\br}' \frac{\dd \bu'}{\Omega_d} \hat{\br}' \left[\left(v_0(\bu' - \bu_1) + \sum_{i>1} \bnabla_{\br_1}V(\br_1-\br_i)\right)\cdot \hat{\br}'\right] \Theta(\hat{\br}'\cdot(\bu_1-\bu'))\hat{K}
\end{split}
\end{equation}
and therefore
\begin{equation}
\begin{split}
& \rho \bnabla_{\br_1}\cdot \int \dd \br' \, \frac{\dd\bu'}{\Omega_d} g^{(q)}(\br_1,...,\br_{q-1}, \br' ; \bu_1,...,\bu_{q-1}, \bu') \bnabla_{\br_1}V(\br_1 - \br') \\ & = -\frac{\rho V_d(\sigma)}{4 d}\bnabla_{\br_1}\cdot\left\{\left[\prod_{i < j} g_0(\br_i,\br_j;\bu_i,\bu_j)\right]\left(-v_0 \bu_1 + \sum_{i>1} \bnabla_{\br_1}V(\br_1-\br_i)\right)\right\}
\end{split}
\end{equation}
up to $O(d^{-1/2})$ corrections. First, this self-consistently justifies that $\frac{\rho V_d(\sigma)}{d}$ is the proper expansion parameter as claimed in Eq.~\eqref{eq:2point-virial}. Second, substituting Eq.\eqref{eq:force} into the hierarchical equation for $g^{(q-1)}$ one can see that the factorized product form of Eq.\eqref{eq:p_point} also holds to order $q-1$. Iteratively, this leads to 
\begin{equation}
g^{(2)}(\br_1,\br_2;\bu_1,\bu_2) = g_0(\br_1,\br_2;\bu_1,\bu_2)
\end{equation}
up to $O(d^{-1/2})$ corrections in this truncated scheme thereby showing that all the functions $g^{(2)}_p$ for $0 < p < q-1$ decay at least as $d^{-1/2}$ as $d \to \infty$. This is valid for any $q > 2$. Thus, upon swapping the summation and the $d \to \infty$ limit, we obtain the central result of this work:
\begin{equation}
g^{(2)}(\br_1,\br_2;\bu_1,\bu_2) = g_0(\br_1,\br_2;\bu_1,\bu_2)
\end{equation}
The factorized product structure of Eq.~\eqref{eq:p_point} for any correlation function is then also proved.

\bibliography{biblio-activemf}